\documentclass[useAMS,usenatbib]{mn2e}

\title[A revised mass model for M31]{A Revised $\Lambda$CDM Mass Model for the Andromeda Galaxy}
\author[M.~S. Seigar et al.]{Marc S. Seigar$^{1, 2, 3}$\thanks{E-mail: mxseigar@ualr.edu (MSS)}, Aaron J. Barth$^2$ and James S. Bullock$^2$\\
$^1$Department of Physics \& Astronomy, University of Arkansas at Little Rock, 2801 S.\ University Avenue, Little Rock, AR 72204, USA\\
$^2$Center for Cosmology, Department of Physics \& Astronomy, University of California, Irvine, 4129 Frederick Reines Hall, Irvine,\\
CA 92697-4575, USA\\
$^3$Arkansas Center for Space and Planetary Sciences, 202 Old Museum Building, University of Arkansas, Fayetteville, AR 72701, USA\\
}
\begin{document}

\date{In original form 1 February 2008}

\pagerange{\pageref{firstpage}--\pageref{lastpage}} \pubyear{2007}

\maketitle

\label{firstpage}

\begin{abstract}

We present an updated mass model for M31 that makes use of a Spitzer 
3.6 $\mu$m image,  a mass-to-light ratio gradient based on the galaxy's $B-R$ 
colour profile, and observed rotation curve data from a variety of sources. We 
examine cases where the dark matter follows a pure NFW profile and where an 
initial NFW halo contracts adiabatically in response to the formation of the 
galaxy. We find that both of these scenarios can produce a reasonable fit to 
the observed rotation curve data. However, a pure NFW model requires a 
concentration $c_{\rm vir}=51$ that is well outside the range predicted in 
$\Lambda$CDM cosmology and is therefore disfavoured. An adiabatically 
contracted NFW halo favors an initial concentration $c_{\rm vir}=20$ and 
virial mass $8.2\times10^{11} M_{\odot}$, and this is in line with the 
cosmological expectations  for a galaxy of the size of M31.
The best-fit mass is consistent 
with published estimates from Andromeda Stream kinematics, satellite galaxy 
radial velocities, and planetary nebulae studies. Finally, using the known 
linear correlation between rotation curve shear and spiral arm pitch angle, 
we show that the stellar spiral arm pitch angle of M31 (which cannot be
deduced from imaging data due to the galaxy's inclination) is 
$P=24\fdg7\pm4\fdg4$.

\end{abstract}

\begin{keywords}
dark matter ---
galaxies: fundamental parameters ---
galaxies: haloes ---
galaxies: individual (M31) ---
galaxies: spiral ---
galaxies: structure
\end{keywords}

\section{Introduction}

Modelling the mass distribution of M31 is a classical problem that has seen
many past iterations (e.g., Einasto 1972; Kent, Huchra \& Stauffer 1989; 
Klypin, Zhoa \& Somerville 2002; Widrow, Perrett \& Suyu 2003; Geehan et al.\
2006). Recently, a Spitzer 3.6 $\mu$m mosaic image of M31 has become available
(Barmby et al.\ 2006). This allows for a more accurate determination of the 
baryonic mass distribution in M31 than has been determined previously. 
Accurate mass modelling of M31 is an important tool for both exploring our 
ideas of cosmology, and for understanding the dynamics of newly discovered 
Local Group dwarf galaxies (e.g., Majewski et al.\ 2007; Chapman et al.\ 2007).

Because M31 can be studied in exquisite detail, it provides a crucial testing 
ground for our ideas of galaxy formation (Kent 1989; Evans \& Wilkinson 2000). 
A problem of particular relevance for Cosmological constant plus Cold Dark 
Matter ($\Lambda$CDM) cosmology is the Tully-Fisher zero-point problem, which 
refers to the fact that standard models cannot reproduce the relation between 
galaxy luminosity  and circular velocity (Tully \& Fisher 1977) without
over-producing the number density of galaxies at fixed luminosity (e.g.\  
Gonzalez et al.\ 2000; Cole et al.\ 2000; Benson et al.\ 2003; Yang, Mo \& 
van den Bosch 2003). Another problem of related concern is the
cusp/concentration problem -- namely that dark-matter dominated galaxy
rotation curves seem to rise more slowly than predicted in $\Lambda$CDM 
(Moore 1994; Flores \& Primack 1994; Moore et al.\ 1999; van den Bosch \& 
Swaters 2001; Blais-Ouellette, Amram \& Carignan 2001; Alam, Bullock \& 
Weinberg 2002; Swaters et al.\ 2003; Simon et al.\ 2005; Dutton et al.\ 2005;
Kuzio de Naray et al.\ 2006). It remains to be seen whether these problems 
point to some unaccounted for process in galaxy formation (e.g.\ Dutton et 
al.\ 2007) or to some new physics of cosmological relevance (Kaplinghat, Knox 
\& Turner 2000; Zentner \& Bullock 2002; Kaplinghat 2005; Cembranos et al.\ 
2005; Strigari, Kaplinghat \& Bullock 2007; Gnedin et al.\ 2007).

Recently, Dutton et al.\ (2007) and Gnedin et al.\ (2007) revisited the
Tully-Fisher problem using two large well-defined samples of disc-dominated
(late-type) galaxies. They both took as a starting point the ``standard'' 
model of disc formation, which assumes that initial Navarro, Frenk \& White 
(1997; hereafter NFW) dark haloes respond to disc formation via adiabatic 
contraction (AC) (Blumenthal et al.\ 1986; Gnedin et al.\ 2004; Sellwood \& 
McGaugh 2005; Choi et al.\ 2006). Both groups conclude that the Tully-Fisher 
zero point cannot be explained if initial NFW haloes have concentrations as 
high as  those expected for $\Lambda$CDM with $\sigma_8 \simeq 0.9$ (with 
$c \sim 12$ for M31-size haloes, e.g.\ Bullock et  al.\ 2001a, Macci\`o et 
al.\ 2006). Both sets of authors agree that the problem could be alleviated 
if AC did not operate (e.g.\ Somerville \& Primack 1999) and Dutton et al.\  
(2007) go on to advocate a model where disc formation induces an 
{\em expansion} in the underlying halo density structure. In contrast, Gnedin 
et al.\ (2007) argue that AC is a fundamental prediction in galaxy formation
and instead suggest that the cosmology be changed to favor lower halo 
concentrations (e.g. Zentner \& Bullock 2002) or that the IMF is lighter 
(with a stellar mass-to-light ratio $M_*/L$ lower by 0.15 dex) than 
the standard Kroupa (2001) assumption.

Given the fundamental issues at hand, the question of whether AC occurs in 
nature is  of significant interest. Theory certainly favours the idea that 
haloes contract. Indeed, halo contraction must occur when the infall of baryons
is smooth and adiabatic (e.g.\  Blumenthal et al.\ 1986; Ryden \& Gunn 1987; 
Sellwood \& McGaugh 2005) and simulations suggest that dark haloes will 
contract even when galaxies or galaxy clusters form quickly from an irregular 
collapse (Barnes 1987; Flores et al.\ 1993; Jesseit, Naab \& Burkert 2002; 
Gnedin et al.\ 2004; Sellwood \&  McGaugh 2005; Choi et al.\ 2006; Weinberg 
et  al.\ 2008). On the other hand, there is very little observational evidence
that halo contraction actually occurs in nature. For example, Zappacosta et 
al.\ (2006) performed a detailed XMM study of the radio-quiet galaxy cluster 
Abell 2589 and conclude that an NFW halo + AC model cannot explain the data 
but that a pure NFW halo provides a remarkable fit down to $\sim 1 \%$ of the 
halo's virial radius. Similarly, an  investigation of seven elliptical 
galaxies with {\em Chandra} by the same group (Humphrey et al.\ 2006) finds 
that AC degrades the mass profile fits significantly unless strong deviations 
from a Kroupa IMF are allowed. Also, Kassin, de Jong \& Pogge (2006a) and 
Kassin, de Jong \& Weiner (2006b) found that the rotation curves of 32 out of 
34 of the bright spiral galaxies they study are better fit without adiabatic 
contraction.

Klypin et al.\ (2002) have shown that the observed rotation curve of M31 is 
best explained with a halo model that includes AC. However, other recent M31 
mass modelling papers (e.g., Widrow et al.\ 2003; Geehan et al.\ 2006) have 
not directly addressed the question of AC. Furthermore, M31 is essentially
the only well-studied galaxy that has been shown to require AC (Klypin et 
al.\ 2002). Most other dynamical studies of galaxies tend to disfavour AC
(e.g., Kassin et al.\ 2006a, b). Given the availability of a 
Spitzer 3.6 $\mu$m image, it is now appropriate to revisit this question
and test the conclusion that M31 requires AC. In 
what follows we show that the rotation curve of M31 can only be explained 
without AC if an unusually high NFW concentration, $c_{\rm vir} \sim 50$, is 
adopted. Such a concentration is well outside the range expected in 
$\Lambda$CDM haloes, and therefore this model is disfavoured. Moreover, the 
best-fit AC model has a slightly high NFW concentration, $c_{\rm vir}=20$, 
for  typical, $\sigma_8= 0.9$, $\Lambda$CDM haloes (Bullock et al.\ 2001a, 
Macci\`o et al.\ 2006), but it is within 1.5$\sigma$ of the expected value 
for an M31 size halo. We speculate on the implications of these results in 
the conclusion section.

Our approach is unique compared to other recently published mass profiles 
derived from modelling the rotation curve of M31. While our method bears most 
similarity to that adopted by Klypin et al.\ (2002), we use an updated 
baryonic mass component, which is derived using a 3.6 $\mu$m Spitzer image. 
Geehan et al.\ (2006) used a pure NFW halo profile, and did not consider AC. 
Tempel et al.\ (2007) also use the 3.6 $\mu$m Spitzer image, yet once again,
AC is not considered.
Lastly, Widrow et al.\ (2003) used a different functional form for the halo 
density profile (see Evans 1993). The Evans profile is based upon an 
isothermal distribution function. Unlike the NFW profile, the Evans profile 
was not derived by fitting models to dark matter haloes produced in 
$\Lambda$CDM simulations. Widrow et al.\ (2003) then go on to fit an NFW model
to their best-fitting Evans (1993) model. 
However, we have chosen to use the cosmologically motivated NFW halo density 
profile model directly.

Throughout this paper we assume a flat $\Lambda$CDM cosmology with 
$\Omega_m=0.27$ and a Hubble constant of $H_0=75$ km s$^{-1}$ Mpc$^{-1}$. We 
relate virial masses ($M_{\rm vir}$) and radii ($R_{\rm vir}$) assuming a 
virial over-density relative to {\em average} of 
$\Delta_{\rm vir}=347$\footnote{Specifically we use $R_{\rm vir}= 206h^{-1} (M_{\rm vir}/10^{12} h^{-1} {\rm M}_{\odot})^{1/3}$ kpc.}.  
At times we quote baryon fractions relative
to the universal value with $f_b=\Omega_b/\Omega_m=0.16$.  
We adopt a distance of 784 kpc to M31 (Holland 1998). As a result, an angular 
distance of $1^{\prime}$ is equivalent to a distance of 228 pc.

\section{Data}
\label{RSS}

We  make  use of the Spitzer/IRAC 3.6 $\mu$m image of M31 (Barmby et al.\ 
2006). The IRAC observations of M31 were taken as part of {\em Spitzer} 
General Observer program 3126 in 2005 August. Fifteen Astronomical Observation
Requests (AORs) were used to map a region approximately $3\fdg7\times1\fdg6$ 
on the sky. The central $1\fdg6\times0\fdg4$ was covered by three AORs, each
having two 12 s frames per position. The outer regions were covered by two 
AORs, each with two dithered 30 s frames per position. The pixel scale of 
IRAC at 3.6 $\mu$m is 1.221 arcsec, with a point spread function (PSF)
that has a FWHM $\sim$1.2 arcsec. 
The 3.6 $\mu$m filter is 1.2 $\mu$m broad, spanning the wavelengths 
2.965--4.165 $\mu$m. For a description of the IRAC instrument see Fazio et 
al.\ (2004). 

The Spitzer data are preferred over 2MASS $K_s$-band data for a few reasons.
The regular 2MASS $K_s$-band image is simply not deep enough to obtain an
accurate bulge/disc decomposition. Furthermore, this same 2MASS image suffers
from over subtraction of the sky by about 0.5 magnitudes (T.\ Jarrett, private
communication). A 6$\times$ 2MASS $K_s$-band image is also available
(where the data were observed with an integration time 6 times longer than
the regular 2MASS survey), but it is only about 1 magnitude deeper and also
hampered by uncertain sky background. The Spitzer image has a larger field of
view, better depth, and a more accurate determination of the sky level.

We also use the $B-R$ colour profile of Walterbos \& Kennicutt (1987), in
order to determine the stellar mass-to-light ratio as a function of radius.
We have corrected this $B-R$ profile has for Galactic foreground extinction,
which is $E(B-V)=0.062$ mags in the direction of M31 (Schlegel, Finkbeiner
\& Davis 1998). This is converted to a correction for the $B-R$ colour by 
using the transformation, $E(B-R)=1.64E(B-V)$ from Schlegel et al.\ (1998),
i.e.\ $E(B-R)=0.102$ for M31. As a comparison the amount of extinction at 
3.6 $\mu$m is 1.5 per cent of that at $B$, and about 34 per cent of that in 
the $K_s$ band.

We adopt the rotation curve data from several sources. In one case (M1) we 
adopt the H$\alpha$ rotation data out to 25 kpc from Rubin \& Ford 
(1970)\footnote{Figure 9 of Rubin \& Ford (1970) also 
includes rotation velocities calculated from a narrow [N{\tt II}] 
$\lambda$6583 emission feature, for the region within 3 kpc. However, these 
velocities are not tabulated in their paper, and so we do not include them 
here.} and 
extend the rotation curve to 35 kpc using H{\tt I} data from Carignan et al.\ 
(2006). In another case (M2) we use the CO rotational velocities from 
Loinard, Allen \& Lequeux (1995) and the H{\tt I} from Brinks \& Burton (1984)
to construct an observed rotation curve out to a 30 kpc radius. This was the 
rotation curve adopted by Klypin et al.\ (2002) in their M31 model. We adopt 
M1 as our fiducial case here because Rubin \& Ford published errors on their 
observed velocities. However, as we show below, both cases yield consistent 
results. In both cases we take into account the presence of a 
$10^{8} M_{\odot}$ supermassive black hole (Bender et al.\ 2005) and the 
H{\tt I} gas distribution (Carignan et al.\ 2006).

\section{Mass modelling of M31}
\label{MM}

\subsection{The baryonic contribution}
\label{bruzual}

Our goal is to determine the best possible mass model for M31. We perform a
bulge-disc decomposition in order to estimate the baryonic contribution to the 
rotation curve. We then determine several possible mass models. The best fit 
model is determined by minimizing the reduced-$\chi^2$ in a fit to the 
observed rotation curve.

\begin{figure}
\includegraphics{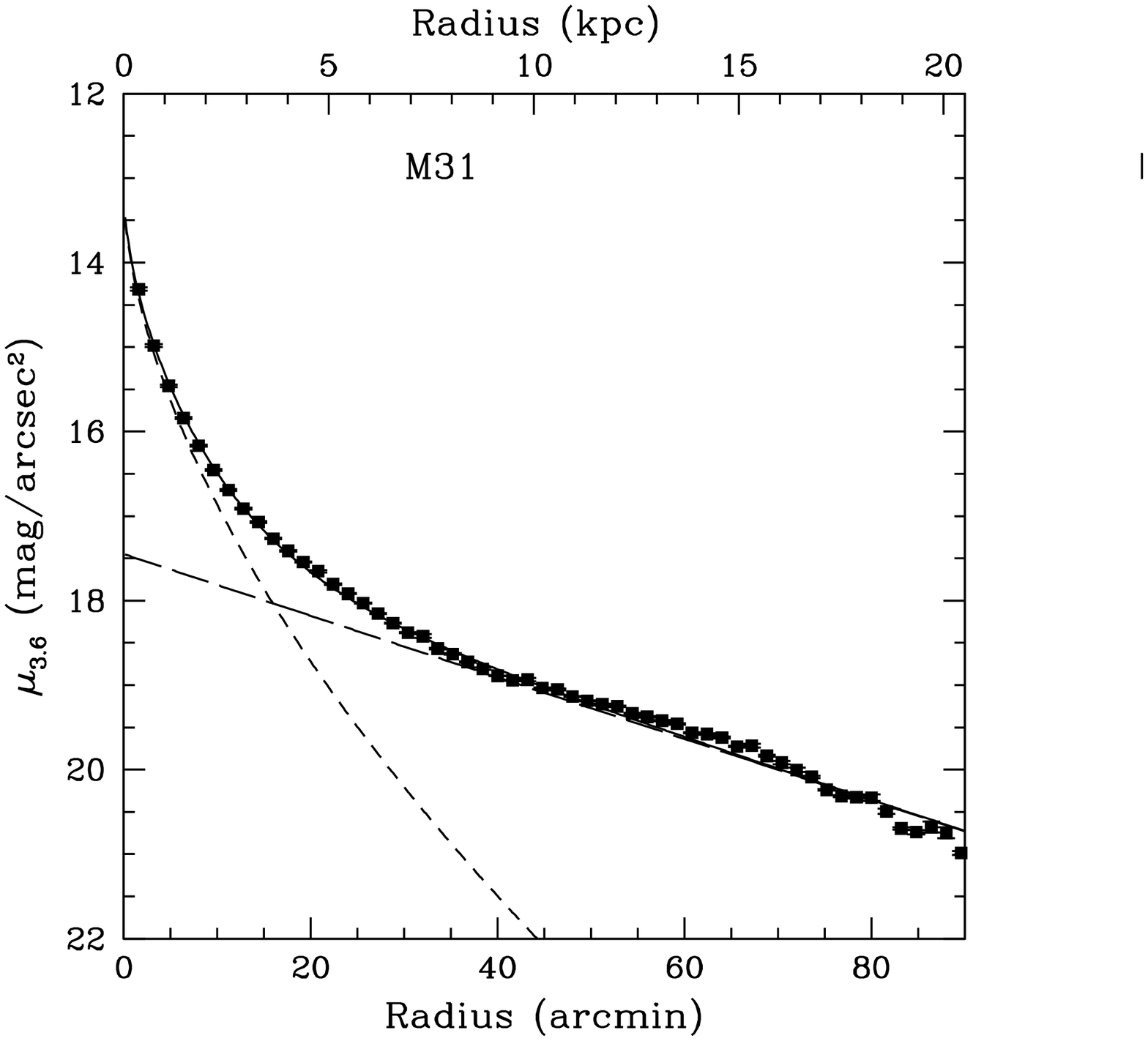}
\vspace*{8.5cm}
\caption{The Spitzer 3.6 $\mu$m surface brightness profile of M31 (without correction for inclination) with decomposition into bulge and disc components. The bulge has been fitted with a S\'ersic model (short dashed line) and the disc has been fitted with an exponential model (long dashed line).}
\label{SB}
\end{figure}

\begin{figure*}
\includegraphics{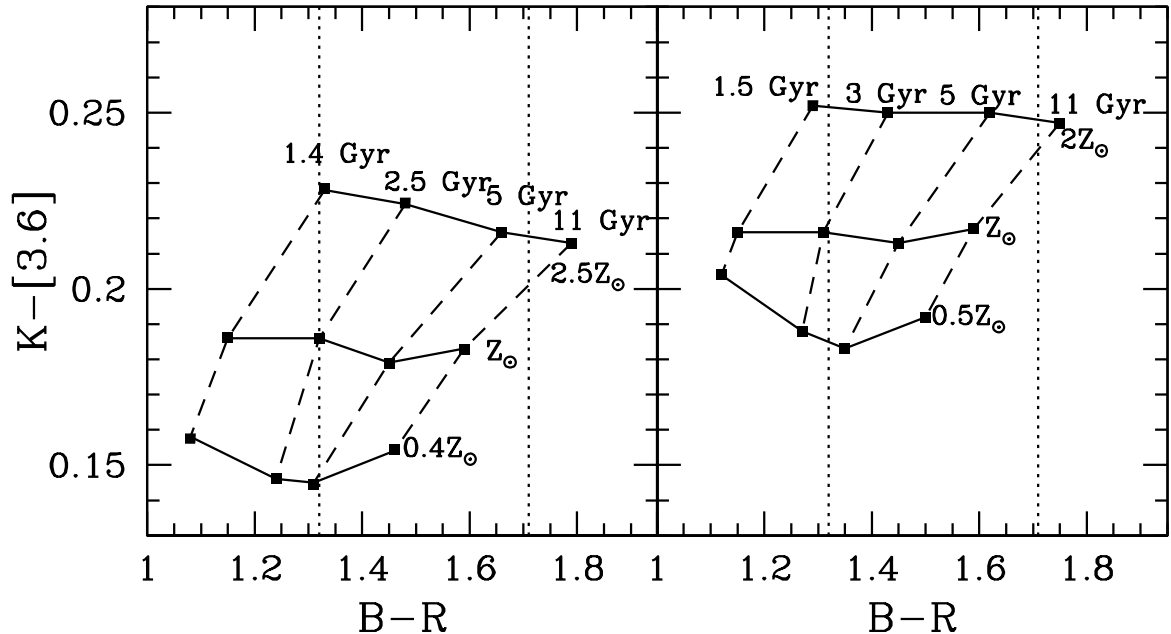}
\vspace*{8.5cm}
\caption{$K-$[3.6] colour as a function of $B-R$ colour for Bruzual \& Charlot (2003) population synthesis models ({\em Left Panel}) and Maraston (2005) population synthesis models ({\em Right Panel}). In both cases, dashed lines represent a constant age stellar population and solid lines represent constant metallicity. The dotted lines represent the typical $B-R$ colour of the inner disc ($B-R=1.79$) and the outer disc ($B-R=1.40$) from Walterbos \& Kennicutt (1987).}
\label{synphot}
\end{figure*}

We first extract the surface brightness profile of M31 using the Spitzer 
3.6 $\mu$m image and the IRAF {\tt ellipse} routine, which fits ellipses to 
an image using an iterative method described by Jedrzejewski (1987). In order 
to mask out foreground stars and satellite galaxies SExtractor (Bertin \& 
Arnouts 1996) was used. An inclination correction was then applied to the 
surface brightness profile (see de Jong 1996; Seigar \& James 1998a) as follows
\begin{equation}
\mu_i = \mu - 2.5C\log{\left(\frac{a}{b}\right)}
\end{equation}
where $\mu_i$ is the surface brightness when viewed at some inclination, $i$,
$\mu$ is the corrected surface brightness, $a$ is the major axis, $b$ is the
minor axis and $C$ is a factor dependent on whether the galaxy is optically
thick or thin; if $C=1$ then the galaxy is optically thin; if $C=0$ then the
galaxy is optically thick. Since we are using infrared data, the correction
for inclination is small, and in many cases completely ignored (see e.g.\
Seigar \& James 1998a; de Jong 1996). However, Graham (2001) showed that 
$C=0.91$ is a good typical value to use for the near-infrared $K_s$-band.
If one adopts a simple reddening law, where extinction falls as the square
of wavelength, then the value of $C=0.97$ is appropriate at 3.6 $\mu$m, and
we adopt this value here.  

The resulting 
surface brightness profile (Figure 1) reaches a surface brightness
of $\mu_{3.6} \sim 21.2$ mag arcsec$^{-2}$ 
at a radius of $1\fdg5$. This is very similar to
the surface brightness found by Barmby et al.\ (2006). Furthermore, the PNe
counts of Merrett et al.\ (2006) predict an R band surface brightness of 
$\sim 23.8$ mag arcsec$^{-2}$. At these radii the metallicity of M31 is 
$0.4 Z_{\odot}$ (Brewer, Richer \& Crabtree 1995; Worthey et al.\ 2005). 
Using the Maraston (2005) population synthesis codes, one can see that the
expected $R-$[3.6] color at this radius is $R-$[3.6]$\sim$2.5, very similar
to our measured surface brightness.

From this surface brightness profile, we perform a one-dimensional bulge-disc 
decomposition, which employs the S\'ersic model for the bulge component and an 
exponential law for the disc component (e.g.\ Andredakis et al.\ 1995; Seigar 
\& James 1998a; Khosroshahi, Wadadekar \& Kembhavi 2000; D'Onofrio 2001; Graham
2001b; M\"ollenhoff \& Heidt 2001; see also Graham \& Driver 2005 for a 
review). The S\'ersic (1963, 1968) $R^{1/n}$ model is most commonly expressed 
as a surface brightness profile, such that
\begin{equation}
\mu(R)=\mu_{e}\exp{\left\{-b_{n}\left[\left(\frac{R}{R_{e}}\right)^{1/n}-1\right]\right\}},
\end{equation}
where $\mu_{e}$ is the surface brightness at the effective radius $R_{e}$ that
encloses half of the total light from the model (Ciotti 1991; Caon, Capaccioli
\& D'Onofrio 1993). The constant $b_{n}$ is defined in terms of the parameter 
$n$, which describes the overall shape of the light profile. When $n=4$ the 
S\'ersic model is equivalent to a de Vaucouleurs (1948, 1959) $R^{1/4}$ model 
and when $n=1$ it is equivalent to an exponential model. The parameter $b_{n}$
has been approximated by $b_{n}=1.9992n-0.3271$, for $0.5<n<10$ (Capaccioli 
1989; Prugniel \& Simien 1997). The exponential model for the disc surface
brightness profile can be written as follows,
\begin{equation}
\mu(R)=\mu_{0}\exp{(-R/h)},
\end{equation}
where $\mu_{0}$ is the disc central surface brightness and $h$ is the disc
scale length. Our bulge-disc decomposition ignores the inner 
$4^{\prime\prime}$ of M31, which is dominated by an independent nuclear 
feature (Light, Danielson \& Schwarschild 1974). The results of the bulge-disc
decomposition can be seen in Figure \ref{SB}. Note that although there appears
to be a slight break in the disc surface brightness profile at 
$\sim$4000$^{\prime\prime}$, the residuals are still very small. This break may
appear due to the presence of spiral structure, evident at similar radii in
the 24 $\mu$m of M31. Observational 
properties of M31 are derived from this bulge-disc decomposition and these 
are listed in Table 1.

\begin{table*}
\label{BDtab}
\caption{M31 Observational Data. Hubble type taken from de Vaucouleurs et al.\ (1991). Distance in kpc taken from Holland et al.\ (1998).}
\begin{tabular}{ll}
\hline
Parameter                          & Measurement   \\
\hline
Hubble type                                           & SA(s)b      \\
Distance (kpc)                                        & 784 \\
Position Angle of major axis (degrees)                & 45   \\
Bulge effective radius, $R_e$ (arcmin)                & 8.45$\pm$0.43 \\
Bulge effective radius, $R_e$ (kpc)                   & 1.93$\pm$0.10 \\
Bulge surface brightness at the effective radius, $\mu_e$ (3.6 $\mu$m-mag arcsec$^{-2}$)  & 16.42$\pm$0.83  \\
Bulge S\'ersic index, $n$              & 1.71$\pm$0.11           \\
Disc central surface brightness, $\mu_0$ (3.6 $\mu$m-mag arcsec$^{-2}$) & 17.35$\pm$0.93  \\
Inclination corrected disc central surface brightness, $\mu_{0_{i}}$ (3.6 $\mu$m-mag arcsec$^{-2}$) & 18.94$\pm$0.93  \\
Disc scalelength, $h$ (arcmin)              & 25.92$\pm$1.28  \\
Disc scalelength, $h$ (kpc)              & 5.91$\pm$0.27 \\
Disc luminosity, $L_{disc}$ ($L_{\odot}$) & (6.08$\pm$0.60)$\times$10$^{10}$ \\
Bulge-to-disc ratio, $B/D$            & 0.57$\pm$0.02          \\
\hline
\end{tabular}
\end{table*}

\begin{figure}
\includegraphics{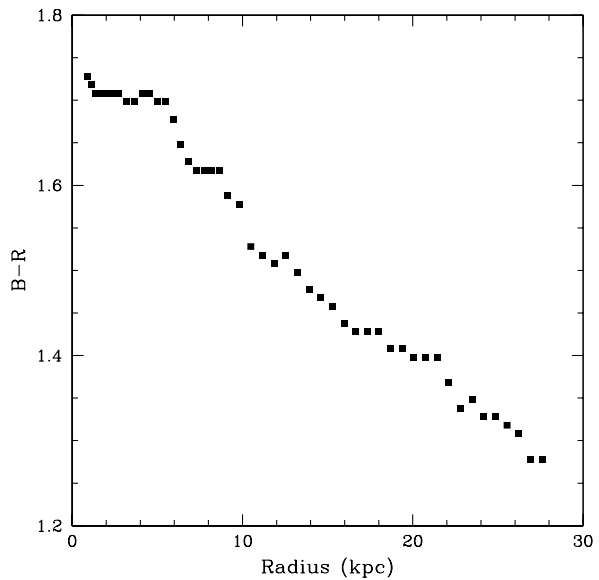}
\vspace*{8cm}
\caption{$B-R$ colour profile for M31 from Walterbos \& Kennicutt (1987) corrected for foreground Galactic extinction.}
\label{colour}
\end{figure}

With a disc scalelength of $25.92^{\prime}$ and a surface brightness
profile extending to a radius of $1\fdg5$, we are tracing the disc profile
out to $\sim$3.5 scalelengths. Assuming that the exponential disc profile
extrapolates to infinity, our profile includes 86.1 per cent of the total disc 
light. Below, where we assign a mass to the disc, we extrapolate the disc
profile to 6 scalelengths, which would contain 98.2 per cent of the disc light.
Since Ibata et 
al.\ (2005) have shown that the disc of M31 extends to very large radii 
($\sim$70 kpc) with the same scalelength, our above calculation assumes no 
disc truncation. Both the 10 kpc ring and the 2.6 kpc 
bar (Beaton et al.\ 2007) show 
up as small perturbations in the surface brightness profile (Fig.\ 1) at 
radii of 46.67$^{\prime}$ and 26.67$^{\prime}$ respectively.

We now assign masses to the disc and bulge of M31. The stellar mass-to-light
ratio in the $K_s$-band is a well calibrated quantity (see Bell et al.\ 2003)
which depends on $B-R$ colour. However, a similar calibration does not exist 
for the Spitzer 3.6 $\mu$m waveband. An alternative method would be to 
estimate the $K_s$-band surface brightness profile, given the 3.6 $\mu$m 
profile, and since we know the $B-R$ colour profile, we have chosen to 
determine the $K-$[3.6] colour dependence on $B-R$ colour using both the 
Bruzual \& 
Charlot (2003; hereafter BC03) stellar population synthesis models and
the Maraston (2005; hereafter M05) stellar population synthesis models.
The BC03
models use the stellar library of Pickles (1998), which includes stars in a 
wide range of spectral types (05-M10) and luminosity classes (I-V) and three 
metallicity groups (metal poor, metal rich and solar). The Pickles (1998) 
library has wavelength coverage from 0.1 $\mu$m to 2.5 $\mu$m. BaSeL 3.1 
spectra (see BC03 Tables 2 and 3) are then used to extend the Pickles (1998) 
spectra into the infrared from 2.5 $\mu$m to 160 $\mu$m. The spectra of M 
giant stars are the only ones not observed in the Pickles (1998) library. As a
result the BC03 models use synthetic M giant spectra calculated by Fluks et 
al.\ (1994). These spectra are extended into the infrared using the model
atmospheres of Schultheis et al.\ (1999), which cover a spectral range from 
0.5 $\mu$m to 10 $\mu$m, and 10 equally spaced stellar temperatures in the 
range $2600<T_{\rm eff}<4400$ K. The main difference between the BC03 and the
M05 models is the way in which thermally-pulsing AGB (TP-AGB) stars are 
treated. The M05 models provide a better treatment of TP-AGB stars than
the BC03 models. Furthermore, Maraston et al.\ (2006) find that 
in general the M05 models
provide better fits to observations of infrared spectra of galaxies
than the BC03 models. They indicate 
systematically lower ages and, on average, masses that are 60 per cent lower 
for the stellar population samples in these galaxies.

Figure \ref{synphot} shows the result of both the BC03 ({\em left panel}) and
M05 ({\em right panel}) models. In both cases, the dotted 
lines in Figure \ref{synphot} represent the $B-R$ colour of the inner and 
outer disc of M31 (Walterbos \& Kennicutt 1987 and Figure \ref{colour}) after
correction for Galactic foreground extinction. Since the M05 models provide
an updated treatment of TP-AGB stars and generally provide better fits to
observations than the BC03 models, from here on we concentrate on the M05
results. However, it is interesting to note that the $K-$[3.6] colour 
index predicted by the BC03 and M05
models differs by $<$0.03 mag (see Figure 2). Lines of constant metallicity 
(solid lines) and age (dashed lines) are shown. The typical metallicity in the
inner few kpc of M31 is $\sim$2.4 $Z_{\odot}$, whereas in the outer disc this 
falls to $\sim$0.4 $Z_{\odot}$ (Brewer, Richer \& Crabtree 1995; Worthey et 
al.\ 2005). From this metallicity gradient and the M05 models (Figure 
\ref{synphot} {\em right panel}) it can be seen that the $B-R$ colour gradient 
in M31 is almost entirely due to changes in metallicity, and that the 
$K-$[3.6] colour difference is $<0.1$ mag. Furthermore, we can now use the 
$B-R$ gradient (Figure \ref{colour}) and the metallicity gradient of M31 to 
estimate the 
$K-$[3.6] colour as a function of radius. We find that the inner region of M31 
has a colour $K-$[3.6]$\simeq 0.25$, and the outer disc (at 26 kpc) has a 
colour $K-$[3.6]$\sim 0.18$. Given the $B-R$ colour profile and the K-band 
mass-to-light ratio calibration from Bell et al. (2003) we calculate a 
central mass-to-light ratio, $M/L_K=1.0\pm0.1$ and a mass-to-light ratio at 
26 kpc, $M/L_K=0.75\pm0.1$. From the $K-$[3.6] colours determined in our 
stellar population synthesis analysis, the above mass-to-light ratios are 
equivalent to 3.6 $\mu$m mass-to-light ratios of $M/L_{3.6}\simeq 1.25\pm0.10$ 
in the centre and $M/L_{3.6}\simeq 0.89\pm0.10$ at 26 kpc (equivalent to a 
$M/L$ gradient 0.13 kpc$^{-1}$). In our analysis, we prefer to keep the 
central $(M/L_{3.6})$ value a free parameter, and as such we allow it to vary 
in the range $0.05<(M/L_{3.6})<1.5$. This is a standard approach to adopt in
mass modelling of galaxies (see Humphrey et al.\ 2006; Zappacosta et al.\ 2006)
and as discussed at the end of section 3.2 and shown in Table 2, when the 
$M/L_{3.6}$ is left as a free parameter, the best-fit
value found is $M/L_{3.6}=1.15$, i.e.\ consistent with the value of $M/L_{3.6}$
derived here. We use our derived value for the 
3.6 $\mu$m disc luminosity of $L_{3.6}=(6.08\pm0.60)\times10^{10}$ $L_{\odot}$.
We then use the derived disc and bulge light profiles 
$L_{3.6}=L_{\rm disc}+L_{\rm bulge}$ to determine the stellar mass 
contribution to the rotation curve, $M_*=(M/L_{3.6})L_{3.6}$.

A concern in using the 3.6 $\mu$m Spitzer waveband to determine the underlying
stellar mass, is the effect of emission from hot dust in this waveband,
although this is probably only important in or near H{\tt II} regions. In
order to place some constraint on this, we have chosen to explore the emission 
from dust in the the near-infared $K$-band at 2.2 $\mu$m. Using near-infrared
spectroscopy at 2.2 $\mu$m it has been shown that hot dust can account for up 
to 30 per cent of the continuum light observed at this wavelength in areas of 
active star formation, i.e., spiral arms (James \& Seigar 1999). When 
averaged over the entire disc of a galaxy, this reduces to a $<$2 per cent 
effect, if one assumes that spiral arms can be up to 12$^{\circ}$ in width. 
At 3.6 $\mu$m this would therefore result in $<$3 per cent of emitted light 
from dust. 

Another concern for the 3.6 $\mu$m waveband, would be the contribution from
the Polycyclic Aromatic Hydrocarbon (PAH) emission feature at 3.3 $\mu$m.
However, ISO spectra covering the wavelength range 5.15 to 16.5 $\mu$m 
have shown that PAH emission in M31 seems to be very 
weak (e.g., Cesarsky et al.\ 
1998). Furthermore, in a survey of actively star forming galaxies, Helou
et al.\ (2000) found that the 3.3 $\mu$m feature was also very weak when
they analysed the average $2.5-11.6$ $\mu$m spectrum of 45 galaxies. The 
contribution of the PAH feature to the 3.6 $\mu$m Spitzer waveband, is 
therefore not a major concern.

In the following section, where we model the dark matter halo, we have included
a bulge surface brightness model using a deprojected S\'ersic function. The
function we use is from Trujillo et al.\ (2002) who show that the deprojected
mass density profile is given by,
\begin{equation}
\rho(R)=\frac{f^{1/2}I(0)b_{n}2^{(n-1)/2n}}{R_{e}n\pi}\Upsilon\frac{h^{p(1/n-1)}K_{\nu}(b_{n}h^{1/n})}{1-C(h)}
\label{dep}
\end{equation}
where $R_e$ is the effective
radius, $I(0)$ is the central intensity, $\Upsilon$ is the stellar 
mass-to-light ratio, $h=R/R_e$, $C(h)=h_1(\log{h})^{2}+h_2\log{h}+h_3$ and 
$K_{\nu}$ is the $\nu^{\rm th}$-order 
modified Bessel function of the third kind. 
Trujillo et al.\ (2002) give values for $\nu$, $p$, $h_1$, $h_2$ and $h_3$ for 
different values of the S\'ersic parameter, $n$. In our case, where $n=1.71$, 
we adopt values of $\nu=0.45394$, $p=0.67930$, $h_1=-0.0162935$, 
$h_2=-0.18161$ and $h_3=0.04435$. Finally, the constant $f^{1/2}$ depends on
the three-dimensional spatial orientation of the bulge. In our case, we 
assume the bulge is intrinsically spherical, and therefore $f^{1/2}=1$.

It should be noted that this version of the deprojected S\'ersic function 
(equation \ref{dep}) is 
an approximation, but it is good enough for our purposes.

\subsection{Modelling the dark matter halo}
\label{halo}

A range of allowed dark matter halo masses and density profiles are now 
explored, using two models for disc galaxy formation. In the first we assume 
that the dark matter halo surrounding M31 has not responded significantly to 
the formation of the disc, i.e., adiabatic contraction does not occur (the 
``no-AC'' model). In this case the dark matter contribution to the rotation 
curve is described by a density profile similar to those found in 
dissipationless dark matter simulations (the NFW profile):
\begin{equation}
\rho(R)=\frac{\rho_{s}}{(R/R_{s})(1+R/R_{s})^{2}}
\end{equation}
where $R_s$ is a characteristic ``inner'' radius, and $\rho_s$ is a 
corresponding inner density. This is a two parameter function and is 
completely specified by choosing two independent parameters, e.g., the virial 
mass $M_{\rm vir}$ (or virial radius $R_{\rm vir}$) and concentration 
$c_{\rm vir}=R_{\rm vir}/R_s$ (see Bullock et al.\ 2001a for a discussion). 
Similarly, given a virial mass $M_{\rm vir}$ and the dark matter circular 
velocity at any radius, the halo concentration $c_{\rm vir}$ is completely 
determined.

In  the second class  of models we adopt  the scenario  of adiabatic
contraction (AC) and specifically use the original scheme discussed by
Blumenthal et al.\ (1986, hereafter B86; see also Bullock et  al.\ 2001b 
and Pizagno et al.\ 2005). We also investigate the slightly revised scheme
presented in Gnedin et al.\ (2004, hereafter G04). As we discuss below, the
G04 scheme reproduces the outer slope of the observed rotation curve of M31, 
but needs a very high concentration to do so.

\begin{figure*}
\includegraphics{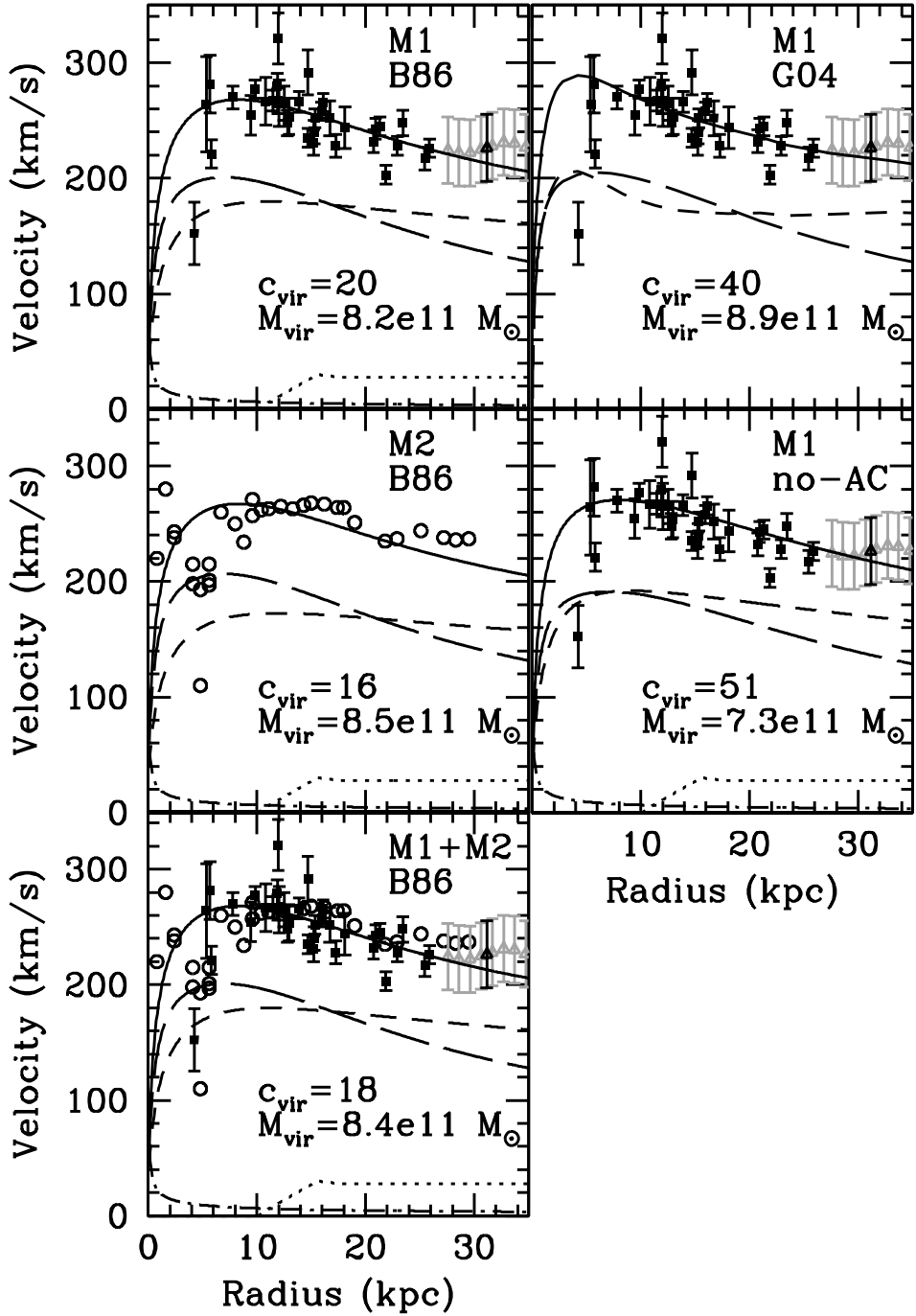}
\vspace*{18.5cm}
\caption{{\em Top left}: Rotation curve data (solid square points) from Rubin \& Ford (1970) with the best fit model, with adiabatic contraction (M1 B86 model), overlaid (solid line). For comparison the H{\tt I} rotation velocities (grey open triangles) from Carignan et al.\ (2006) are shown (the black point corresponds to the average velocity and radius of the Carignan data that we make use of). The best fit rotation curve is decomposed into four components, the contribution from dark matter (short-dashed line), the contribution from baryonic matter (long-dashed line), the H{\tt I} contribution (dotted line) from Carignan et al.\ (2006) and the contribution from the central $10^{8}M_{\odot}$ black hole (dot-dashed line) from Bender et al.\ (2005). {\em Middle left}: Rotation curve CO and H{\tt I} data (solid square points) from Loinard et al.\ (1995) and Brinks \& Burton (1984) respectively, as used by Klypin et al.\ (2002). The best fit model with adiabatic contraction (M2 B86 model) is overlaid (solid line). It is decomposed into four components as in the top left panel. 
{\em Bottom left}: Rotation curve data from Rubin \& Ford (1970; solid points) and Carignan et al.\ (2006; triangles) and the data as used by Klypin et al.\ (2002; hollow circles), with adiabatic contraction model (M1+M2 B86), overlaid.
{\em Top right}: The same as the top left panel, but with the best fitting model using a modified version of adiabatic contraction (M1 G04 model) as in Gnedin et al.\ (2004). The long-dashed curve includes the total baryonic mass, including the H{\tt I} mass and the central black hole. {\em Middle right}: The same as the top left panel, but with the best fitting model without an adiabatically contracted halo (M1 no-AC model). In all five cases the NFW concentration, $c_{\rm vir}$, and virial mass, $M_{\rm vir}$, are shown.}
\label{rcurve}
\end{figure*}

\begin{table*}
\label{modTab}
\caption{M31 best fitting models. ``M1 B86'' is the best-fit model to the H$\alpha$ rotation curve from Rubin \& Ford (1970) (M1) using the AC prescription of B86. ``M2 B86'' is the best-fit B86 model to a combination of the CO rotation velocities from Loinard et al.\ (1995) and the H{\tt I} velocities from Brinks \& Burton (1984) (M2). ``M1+M2 B86'' is the best-fit B86 model to the H$\alpha$ rotation curve from Rubin \& Ford (1970) (M1) combined with the CO rotation velocities from Loinard et al.\ (1995) and the H{\tt I} velocities from Brinks \& Burton (1984) (M2). ``M1 G04'' corresponds to the M1 data fit using the modified adiabatic contraction model of Gnedin et al.\ (2004) and ``M1 no-AC'', uses the same data without adiabatic contraction.}
\begin{tabular}{lccccc}
\hline
Parameter       & M1     & M2     & M1+M2  & M1     & M1     \\
                & B86    & B86    & B86    & G04    & no-AC  \\
\hline
Shear                                                          & 0.51                & 0.50           & 0.51           & 0.54           & 0.50 \\
\\
$V_{2.2}$ (km s$^{-1}$)                                        & 275$\pm$5           & 275$\pm$5      & 275$\pm$5      & 275$\pm$5      & 275$\pm$5 \\
\\
Disc luminosity, $L_{disc}$ ($L_{\odot}$)                      & $(4.9\pm0.6)\times 10^{10}$ & $(6.1\pm0.6)\times 10^{10}$ & $(5.5\pm0.6)\times 10^{10}$ & $(4.9\pm0.6)\times 10^{10}$ & $(6.1\pm0.6)\times 10^{10}$\\
\\
Disc scale length, $h$ (kpc)                                   & 5.09$\pm$0.06       & 5.91$\pm$0.06           & 5.50$\pm$0.06  & 5.50$\pm$0.06  & 5.91$\pm$0.06 \\
\\
Central disc mass-to-light ratio,                              & 1.15$\pm$0.10       & 1.00$\pm$0.10    & 1.05$\pm$0.10   & 1.05$\pm$0.10  & 0.95$\pm$0.10 \\
$M/L$ (3.6 $\mu$m solar units)\\
\\
Mass-to-light ratio gradient,                                  & 0.013               & 0.013          & 0.013          & 0.013          & 0.013         \\
$d(M/L)/dR$ (kpc$^{-1}$)\\
\\
Initial NFW concentration, $c_{\rm vir}$                       & 20.0$\pm$1.1        & 16.0$\pm$1.1   & 18.0$\pm$1.1   & 40.0$\pm$1.1   & 51.0$\pm$1.1   \\
\\
Bulge-to-disc ratio, $B/D$                                     & 0.57$\pm$0.02       & 0.57$\pm$0.02  & 0.57$\pm$0.02  & 0.57$\pm$0.02  & 0.57$\pm$0.02 \\
\\
Virial mass, $M_{\rm vir}$ ($M_{\odot}$)                       & $(8.2\pm0.2)\times10^{11}$  & $(8.5\pm0.2)\times10^{11}$ & $(8.4\pm0.2)\times10^{11}$ & $(8.9\pm0.2)\times10^{11}$ & $(7.3\pm0.2)\times10^{11}$\\
\\
Dark matter concentration,                                     & 14.9$\pm$1.4\%      & 14.4$\pm$1.5\% & 14.7$\pm$1.5\% & 9.9$\pm$1.3\%  & 14.1$\pm$2.0\%   \\
$c_{\rm DM}$, $R<10$ kpc \\
\\
Total mass concentration,                                      & 26.2$\pm$2.5\%      & 25.4$\pm$2.4\% & 25.9$\pm$2.4\% & 23.8$\pm$2.7\% & 26.5$\pm$2.6\%  \\
$c_{\rm tot}$, $R<10$ kpc\\
\\
Stellar baryon fraction, $f_{*}$                               & 56.7$\pm$3.9\%      & 59.2$\pm$4.5\% & 58.1$\pm$4.2\% & 56.3$\pm$6.0\% & 63.5$\pm$5.6\%        \\
\\
$\chi^2$                                                       & 38.75               & 40.63          & 80.31      & 46.50          & 41.58  \\
\\
Degrees of freedom, $\nu$                                      & 31                  & 27             & 64         & 31             & 31    \\
\hline
\end{tabular}
\end{table*}

For each AC algorithm (B86 and G04) and for the no-AC model we generate a grid
of final rotation curves.  We vary the baryonic contribution to the rotation 
curve by allowing the  bulge-disc decomposition parameters for $h$ and 
$L_{\rm disc}$ to range over their $\pm 2\sigma$ values (see Table 1 and the 
previous section). The central mass-to-light ratio of the galaxy is allowed 
to vary over the range $0.05<(M/L)<1.5$. From this we determine the bulge and 
disc contribution to the rotation curve. For the bulge a simplified spherical
model is used. For the disc the potential is derived assuming zero thickness 
and modified Bessel functions as described by Binney \& Tremaine (1987). For 
each set of baryonic parameters we generate a range of dark halo models with 
total circular velocities at $2.2$ disc scalelength ($V_{2.2}$) that span the
best-fit value for M31 ($V_{2.2}=270$ km s$^{-1}$) quoted by Rubin \& Ford 
(1970) by a wide margin:  $200<V_{2.2}<340$ km s$^{-1}$. In practice we 
achieve each $V_{2.2}$ value in our models by allowing the initial halo 
concentration to vary over $>4\sigma$ of its expected range $c_{\rm vir}=3-60$
(Bullock et al.\ 2001a; Wechsler et al.\ 2002; Macci\`o et al.\ 2006) and then
setting the halo virial mass $M_{\rm vir}$ necessary to reproduce the desired 
value of $V_{2.2}$. Finally, we determine the implied fraction of the mass in 
the system in the form of stars compared to that expected from the Universal 
baryon fraction, $f_*=M_*/(f_{b}M_{\rm vir})$ and demand that $f_*$ obeys 
$0.01f_b<f_*<f_b$. From this grid of choices we derive best-fitting dark halo
parameters by minimizing the reduced-$\chi^2$ for our two choices of rotation 
curve data. These results are summarized in Table 2.

For the most part, when minimizing $\chi^2$, we assume that the 
errors on the observational data (i.e., the rotation velocities) are not 
correlated. Instead we assume that they are Gaussian in nature. However, this 
is  not the case for the Carignan et al.\ (2006) H{\tt I} data (see 
Figure \ref{rcurve} top row and bottom right beyond 25 kpc), where the errors 
obviously are correlated. In order to account for the strange behaviour of the
errors on these H{\tt I} rotation velocities we have treated all of this data 
as an average rotation velocity of 226.3 km s$^{-1}$ 
at 31.16 kpc. This is an idealized 
simplification but allows a quantitative estimate of the relative goodness of 
fit between models. Furthermore, although error bars are not available for 
the ``M2'' curve, we assume 10 per cent errors in the model fitting, as this 
is similar to the size of the errors quoted for the Rubin \& Ford (1970) 
H$\alpha$ rotation curve. We therefore quote a $\chi^2$ for 
the ``M2'' curve in Table 2 that is consistent with 10 per cent errors. 
We then determine how consistent all the derived 
rotation curves are with each other, by adopting the formal best fitting 
rotation curve as our fiducial mass model. For our ``M1'' curve, the number 
of degrees of freedom in the fit is $\nu=31$.  The number of data points we
fit to is $ n_{\rm data}=37$, and 6 parameters are determined from the fit,
the NFW concentration $c_{\rm vir}$, the halo virial mass $ M_{\rm vir}$,
the stellar mass-to-light ratio $M/L$, the disc luminosity $L_{\rm disc}$,
the disc scalelength $h$ and the bulge-to-disc ratio $B/D$, thus leaving
the degrees of freedom in the fit as $\nu=31$.

The M1 rotation curve from Rubin \& Ford (1970) is shown by the solid squares 
with error bars in Figure \ref{rcurve} ({\em top left, top right, middle 
right, bottom left}) 
along with our best-fitting rotation curve models (solid lines). 
The contribution from stellar mass (short-dashed) and dark matter 
(long-dash) in each case is shown along with the rather minor contributions 
from neutral gas (dotted) and the central black hole (dot-dashed).

Overlaid in the {\em top left} panel of Figure \ref{rcurve} is a model that 
includes adiabatic contraction (overall reduced-$\chi^2=1.25$). The dark halo
in this case {\em has undergone} adiabatic contraction according to the B86 
prescription. Our best-fitting parameters are summarized in Table 2 under 
``M1 B86''.  We determine the halo of M31 to have a total (dark+baryonic) 
virial mass of $M_{\rm vir}=(8.2\pm0.2)\times10^{11}M_{\odot}$ and an initial 
concentration $c_{\rm vir}=20.0\pm1.1$. Our derived virial mass is almost a 
factor of 2 smaller than that quoted by Klypin et al.\ (2002). We also find
a concentration that is higher than both Klypin et al.\ (2002), who find
$c_{\rm vir}=12$ and Widrow et al.\ (2003), who find $c_{\rm vir}=10$. In
the case of Widrow et al.\ (2003) it should be noted that they initially fit
an isothermal model given by Evans (1993). They then fit an NFW profile
to this model. Given that the isothermal model favors a constant density core,
the derived NFW concentration will be articifially low and fitting an NFW
model in a more direct manner may result in a higher concentration.

The model rotation curve fits the data adequately within the uncertainties. 
The innermost 
data point from the observational rotation curve is significantly lower than 
the model. This can be explained by the recent strong evidence for a bar, 
which dominates the near-infrared light out to a radius of $\sim$2.6 kpc, in 
M31 (Athanassoula \& Beaton 2006; Beaton et al.\ 2007). This bar has been used
to explain the minimum seen in the observed rotation curve at a radius of 
$\sim2.7$ kpc (Athanassoula \& Beaton 2006). This bar may also affect the 
outer rotation velocities as well, although probably by less than 10 per cent.
This would therefore be within the errors associated with the observed data. 
Indeed, shifting the outer rotation velocities systematically upward by 
$\sim$10 per cent changes the best-fit models by an amount that remains within
the quoted error range of the original fit (see Table 2). It should 
also be noted that by a radius of $\sim26-35$ kpc the rotational velocity has 
dropped to $\sim 225-230$ km s$^{-1}$, which is similar to the rotational 
velocities seen at radii of $25-35$ kpc in the H{\tt I} rotation curve 
(Carignan et al.\ 2006; open triangles).

\begin{figure}
\includegraphics{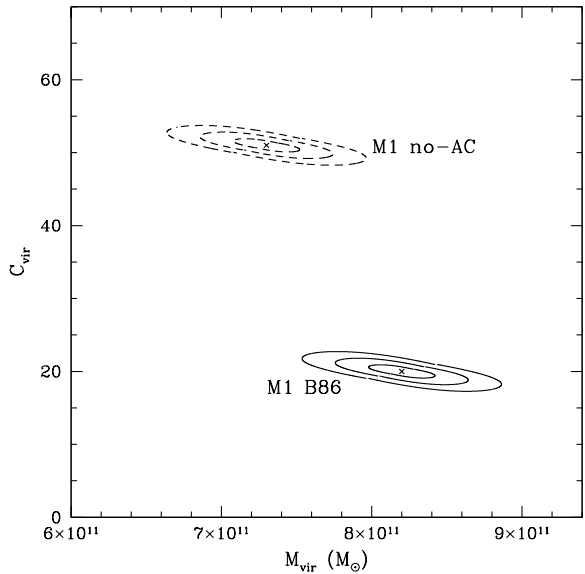}
\vspace*{8cm}
\caption{Virial mass as a function of NFW concentration from our modelling with our 1$\sigma$, 2$\sigma$ and 3$\sigma$ confidence contours shown as solid lines for our ``M1 B86'' model and as dashed lines for our ``M1 no-AC''.}
\label{chi2}
\end{figure}

The  best-fitting model without adiabatic contraction  is shown by the solid 
line in the {\em middle  right} panel of Figure \ref{rcurve}. This, the no-AC
model, was determined in exactly  the same way as our best-fit B86 AC model.  
As listed in Table 2, the no-AC assumption requires a much higher NFW 
concentration, $c_{\rm vir}=51.0\pm1.1$ and a lower virial mass, 
$M_{\rm vir}=(7.3\pm0.2)\times10^{11}M_{\odot}$. This model also produces a 
reasonable fit, with a reduced-$\chi^2=1.35$, but it should be noted that the 
required concentration is, at best, only marginally acceptable within a 
$\Lambda$CDM context. This would correspond to a $\sim 4.4 \sigma$ outlier 
from the expected concentration distribution of M31-sized haloes (e.g., Bullock
et al.\ 2001a).

In Figure \ref{chi2} we plot the NFW concentration, $c_{\rm vir}$, as a 
function of the virial mass, $M_{\rm vir}$, for both the ``M1 B86'' case 
({\em solid line}) and the ``M1 no-AC'' case ({\em dashed line}). The contours
represent the 1$\sigma$, 2$\sigma$ and 3$\sigma$ 
confidence levels for the concentration and virial mass for both models.
From figure \ref{chi2}, one can see that the ``M1 no-AC'' model
is inconsistent with any reasonable concentration of $c_{\rm vir}<30$ at the 
5$\sigma$ level at least. However, since it is 
possible to produce a reasonable fit with a pure NFW model, albeit with a 
very large concentration, we do not rule out our ``M1 no-AC'' case, although 
it is clearly not consistent with standard $\Lambda$CDM cosmology. 

The confidence contours shown in Figure \ref{chi2} have also allowed us to 
estimate errors on the derived best-fit masses and concentrations.
The 1$\sigma$, 2$\sigma$ and 3$\sigma$ 
contours are drawn as the boundaries where 
$\chi^2$ increases by 1.0, 2.71 and 6.63 relative to its minimum value. 
For our best-fit `M1 B86' model, $\chi^2$=38.75. There is also a `M1 B86'
model for which the $\chi^2$=39.75, i.e.\ 1$\sigma$ higher than the best fit. 
For this model, the concentration is $c_{\rm vir}$=21.1 and the virial mass is 
$M_{\rm vir}=8.0\times10^{11} M_{\odot}$. There is a change in the 
concentration
of $\Delta c_{\rm vir}$=1.1 and a change in the virial mass of 
$\Delta M_{\rm vir}=0.2\times10^{11} M_{\odot}$ at the 1$\sigma$ level, i.e.
when the value of $\chi^2$ is increased by 1. The values of
$\Delta c_{\rm vir}$ and $\Delta M_{\rm vir}$ are therefore adopted as our
1$\sigma$ error on $c_{\rm vir}$ and $M_{\rm vir}$ respectively, and these
are the errors that we report in Table 2.

An intermediate result is shown in the {\em top right} panel of Figure 
\ref{rcurve}. Here we use the AC model of G04, which produces less contraction
than the B86 prescription. If the initial NFW concentration is held fixed at 
the same value as the best-fitting B86 case, i.e., $c_{\rm vir}=20$, then the 
G04 model overestimates the observed rotation curve at $r>15$ kpc. The G04 
model can only reproduce this outer slope when the initial halo has 
$c_{\rm vir}=40$. With this high concentration the reduced-$\chi^2$ of the 
G04 model is 1.50. This model cannot be ruled out, but (as in the pure NFW 
case) its halo concentration is higher than expected in $\Lambda$CDM (Bullock 
et al.\ 2001a, Macci\`o et al.\ 2006). Therefore, from here on, we chose to 
focus on the B86 AC case, but provide the best-fitting model results for the 
G04 case in Table 2.

For completeness, we also perform a fit to the same observed rotation curve 
that was used by Klypin et al.\ (2002) in their M31 mass decomposition. Our 
best fitting mass model is shown in the {\em middle left} panel of Figure 
\ref{rcurve} (``M2 B86'' in Table 2). Here we have adopted the B86 AC model.
The initial NFW concentration in this case ($c_{\rm vir}=18$) is higher than 
that the value of $c_{\rm vir}=12$ favored by Klypin et al.\ (2002) and lower 
than our ``M1 B86'' case. However, it is not a large difference, and at face
value our ``M1 B86'' and ``M2 B86'' models are consistent with each other. As 
we discuss below, the difference between our result and that of Klypin et al. 
(2002) is driven mainly by our updated baryonic model, which seems to
provide a less concentrated bulge. For the M2 case, we 
calculate a $\chi^2$, assuming 10 per cent errors. As a final consistency
check, we also perform a fit of the B86 AC model to a combination of the M1
and M2 data ({\em bottom left} of Figure \ref{rcurve}; ``M1+M2 B86'' in 
Table 2). This produces a result consistent with ``M1 B86'' and ``M2 B86''.

\begin{figure}
\includegraphics{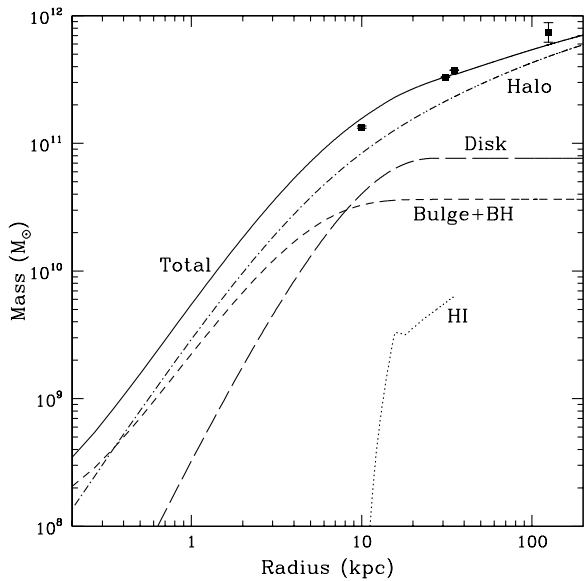}
\vspace*{8cm}
\caption{Enclosed mass as a function of radius for model M1 B86. The solid line indicates the total mass. Also indicated are the HI mass out to a 35 kpc radius (dotted line), the bulge+BH mass (short-dashed line), the disc mass (long-dashed line) and the halo mass (dot-dashed line). The data points indicate masses derived using other observational methods at fixed radii of 10 kpc, 31 kpc, 35 kpc and 125 kpc from Rubin \& Ford (1970), Evans \& Wilkinson (2000), Carignan et al.\ (2006) and Fardal et al.\ (2006) respectively.}
\label{mass}
\end{figure}

Figure \ref{mass} shows the enclosed mass as a function of radius for our 
fiducial model (M1 B86). The fraction of mass contained within the central 
10 kpc (the ``central mass concentration'' hereafter, as originally defined by
Seigar et al.\ 2006) is $c_{\rm tot}=26.2\pm2.5$ per cent. In absolute terms, 
the mass contained within 10 kpc is $(21.5\pm2.6)\times10^{10} M_{\odot}$.  
This is about a factor of 2 larger than that calculated by Rubin \& Ford (1970),
but it should be noted that they did not take into account non-circular orbits
due to the presence of a bar, which we now know exists in M31. This would have
the effect of increasing the mass contained within the bar region. The mass 
calculated within a 35 kpc radius from the H{\tt I} rotation curve by Carignan
et al.\ (2006) is $3.4 \times 10^{11} M_{\odot}$, similar to the mass derived
from our best fit model within a 35 kpc radius of 
$(3.4\pm0.5)\times10^{11} M_{\odot}$. Within a 31 kpc radius we find a mass 
of $(3.1\pm0.4) \times 10^{11} M_{\odot}$, which is similar to the mass of 
$2.8 \times 10^{11} M_{\odot}$ within the same radius found using kinematic 
data of planetary nebulae (Evans \& Wilkinson 2000). The total implied stellar
baryon fraction of this model is $f_{*}=56.7\pm3.9$ per cent.

Other mass models of M31 (Klypin et al.\ 2002; Geehan et al.\ 2006; Widrow et 
al.\ 2003) have used bulge velocity dispersions as inputs into their models.
For example, Klypin et al.\ (2002) used the line-of-sight velocity dispersion 
of M31 from Kormendy \& Bender (1999) to constrain mass models of M31 within 
the bulge region. In the bulge region, the velocity dispersion, $\sigma$ is 
dominant (c.\ 150 km s$^{-1}$) at least out to twice the effective radius, 
i.e.\ $2R_e=$3.96 kpc. Geehan et al.\ (2006) assume a Hernquist 
(1990) model for the bulge, and compare the expected velocity dispersion
profile from Jeans equation modelling with the observations of Kormendy (1988).
Our approach differs from this. We have chosen to assume a S\'ersic law
for the Bulge and we then perform a mass-to-light conversion. In our approach,
we essentially ignore any kinematics within the inner few kpc, because in that
region the kinematics are likely to be perturbed by the bar or affected by
streaming motions along the bar. Instead, we assume
that stellar light traces stellar mass. Figure 4 shows that the
bar kinematics are strongest within the inner $\sim$6 kpc. Beyond 10
kpc, the bar kinematics are not likely to affect the disc kinematics.  
Furthermore, we assume that the bulge is spherically symmetric (an assumption 
also made in the papers by Klypin, Geehan and Widrow). The bulge in our model 
has a mass of $M_{bulge}\simeq3.5\times10^{10}$ $M_{\odot}$. It is reassuring 
to note that the Jeans modelling of Geehan et al.\ (2006) derives a bulge mass 
of $M_{bulge}=3.2\times10^{10}$ $M_{\odot}$, very similar to our bulge mass.

We note that the bulge component in our models is less concentrated than
that derived by some other groups. Klypin et al.\ (2002) have a more 
concentrated bulge, but that is probably due to a combination of the fact that 
they use a Hernquist model (equivalent to a S\'ersic index of $n=4$) for the 
M31 bulge and optical data to determine the bulge characteristics. Geehan et 
al.\ (2006) also use a Hernquist model. This type of model will naturally 
produce a more concentrated bulge than a S\'ersic model with index 
$n\simeq1.7$. The use of the $M/L$ ratios from Bell \& de Jong (2001) also 
gives Klypin et al.\ (2002) an overall more massive bulge. The more recent 
models by Tamm et al.\ (2007) and Tempel et al.\ (2007) make use of a model 
that is similar to a S\'ersic law, but they find an index of $n\simeq3$ 
compared to our index of $n\simeq1.7$. As a result, their bulge model is also 
naturally more concentrated than ours. It is also interesting to note that
replacing our S\'ersic bulge with a bulge profile modelled using a 
Hernquist or exponential law does not affect the overall dark matter halo
profile by more than 10 per cent. This may be because the overall bulge
mass does not change; It is just the way in which the bulge mass is
distrbuted that changes.

The issue of turbulent gas motions is very important for low-mass galaxies
with small rotation velocities (e.g., Valenzuela et al.\ 2007), as this 
requires a correction for pressure support or asymmetric drift in the mass 
model. The correction to the total mass is 
$\sim(\sigma_{\rm gas}/V_{\rm rot})^2$.
Since the H$\alpha$ and H{\tt I} rotation curves are measured from gas, it is 
important to understand what effect the turbulent gas motions may have on our 
mass model. Since, M31 is a galaxy with a high rotation velocity ($\sim 275$ km
s$^{-1}$) and a turbulent velocity dispersion in galaxy discs is typically
$<30$ km s$^{-1}$ (i.e.\ around 10 per cent of the rotation velocity) this 
effect is small. Indeed, if we assume that the scatter of the H$\alpha$ Rubin
\& Ford (1970) rotation curve around our best fitting model is due to turbulent
gas motions, then we can derive an estimate for their typical value in M31.
We can do this because the Rubin \& Ford (1970) rotation curve is derived from
H$\alpha$ spectroscopic observations of individual H{\tt II} regions in M31. 
In this case, we have to assume that the errors on the Rubin \& Ford (1970) 
rotation curve are uncorrelated, and we derive a turbulent velocity dispersion
of $\sigma_{\rm gas}=18.3$ km s$^{-1}$. Adding this in quadrature to the 
maximum rotation velocity from the Rubin \& Ford (1970), results in a mass 
difference of $\sim$0.4 per cent for the {\em total} (baryonic + halo) galaxy 
mass.


As mentioned above, the preferred total (dark+baryonic) virial mass for M31 is 
$(8.2\pm0.2) \times 10^{11} M_{\odot}$. This is similar to the estimate of
$\sim (7.9\pm0.5) \times 10^{11} M_{\odot}$ derived from satellite kinematics
(C\^ot\'e et al.\ 2000) and the total mass of $\sim 8 \times 10^{11} M_{\odot}$
from kinematics of the Andromeda Stream (Fardal et al.\ 2006; Geehan et al.\ 
2006). It is also close to the lower limit of $9\times10^{11}M_{\odot}$ 
derived from kinematics of halo stars (Chapman et al.\ 2006). Our favored 
virial mass is about a factor of 2 smaller than the value favored by the 
Klypin et al.\ (2002) analysis of the M31 rotation curve. This can be 
attributed to three differences in our analysis. First, Klypin et al.\ (2002) 
use the H{\tt I} rotation curve of Brinks \& Burton (1984) to find their best 
fitting model. In this study we use the H$\alpha$ rotation velocities of Rubin
\& Ford (1970). However, we find that this is not the major driver. We find 
that the model rotation velocities are consistent with those from the more 
recent H{\tt I} study of Carignan et al.\ (2006), and when we fit to the same 
observed rotation curve as Klypin et al.\ (2002), we find that our best-fit 
model is consistent with our best-fit model to the Rubin \& Ford (1970) 
curve. Furthermore, if we try and model a combination of data from Rubin
\& Ford (1970) and Klypin et al.\ (2002) we also find a best-fitting rotation
curve consistent with our fiducial case.
More importantly, Klypin et al.\ (2002) used the $M/L$ ratios of Bell 
\& de Jong (2002) to determine the stellar mass from their bulge-disc 
decomposition. In this study we allow $M/L$ to vary freely in the range
$0.05<(M/L)<1.50$, and find a best-fitting value close to the expected
3.6 $\mu$m mass-to-light ratios (see section \ref{bruzual}). Possibly the most 
significant difference is that Klypin et al.\ (2002) perform a bulge-disc 
decomposition using an $R$ band image, whereas we use a Spitzer 3.6 $\mu$m 
image. Using a 3.6 $\mu$m image provides a better trace of the stellar mass 
(e.g., Barmby et al.\ 2006 and references therein). Overall, we should 
therefore have a more accurate estimate of the baryonic mass profile in M31. 
Indeed, if we adopt the baryonic mass profile of Klypin et al.\ (2002) and 
try to fit either to the H{\tt I} or to the H$\alpha$ rotation curve, we find 
a best fitting model that is consistent with their result. 

Typically, the best-fitting $M/L$ values
we find are in the range 0.95--1.15, depending on
the model used. The lowest value of $M/L=0.95$ is difficult to reconcile
with the expected value, which is $M/L_{3.6}=1.2\pm0.1$, for a 
galaxy with the central $B-R$ colour of M31. This expected value was
determined by combining the Bell et al.\ (2003) $K_s$-band $M/L$ with the 
$K-$[3.6] colours derived from our stellar population synthesis described in 
section 3.1 (Figure \ref{synphot}). The lowest value we find ($M/L=0.95$) is
for our no-AC model. We also find that this model is difficult to explain
in the context of $\Lambda$CDM cosmology, due to its large concentration.
Our M1 B86 model reproduced the highest value of $M/L=1.15$ that we find.
This is consistent with the expected range of $M/L$ values.

\section{Summary}

In this paper we have derived new mass models for M31, based on an updated 
baryonic contribution to the rotation curve. The baryonic contribution is 
derived using a Spitzer 3.6 $\mu$m image. We find that both a pure NFW and an 
adiabatically contracted NFW profile can produce reasonable fits to the 
observed rotation curve of M31. However, the best-fit concentration of M31 for
a pure NFW model is $c_{\rm vir}=51\pm1.1$, which is disfavoured when compared 
to the expected value of $\log_{10}(c_{\rm vir})=1.08\pm0.14$ (Wechsler et 
al.\ 2002; Maccio et al.\ 2006) in a $\Lambda$CDM cosmology (our best fitting 
concentration is a $4.4\sigma$ outlier). The best fit produced by the AC 
recipe of Gnedin et al.\ (2004) brings down the concentration to 
$c_{\rm vir}=40$, or $\log_{10}(c_{\rm vir})=1.60$, and is thus marginally 
consistent with $\Lambda$CDM as a $\sim$3.7$\sigma$ outlier. If one accepts 
$\Lambda$CDM cosmology, then the most consistent fit to the observed rotation 
curve seems to be that produced by the AC recipe of Blumenthal et al.\ (1986).
In this case the concentration, $c_{\rm vir}=20\pm1.1$, and this is consistent
with the expectations of $\Lambda$CDM, to within 1.5$\sigma$.

For all of the types of model we use, the derived best-fits are
in good agreement with the mass distributions derived from various other 
observational methods. Our estimate of the halo virial mass lies in the
range
$M_{\rm vir}=(7.3\pm0.2)\times10^{11} M_{\odot}$ (for the M1 no-AC case) 
and $M_{\rm vir}=(8.9\pm0.2)\times10^{11} M_{\odot}$ (for the M1 G04 case).
This is in close agreement with 
the virial mass of $\sim 8\times 10^{11} M_{\odot}$ found via kinematics of 
satellite galaxies (C\^ot\'e et al.\ 2000) and the Andromeda Stream (Fardal 
et al.\ 2006; Geehan et al.\ 2006), and the lower limit of 
$9\times10^{11} M_{\odot}$ found from kinematics of halo stars (Chapman et 
al.\ 2006). Our M1 B86 model, which is in closest agreement with the
expectations of $\Lambda$CDM, lies in the middle of the range with a virial
mass of $M_{\rm vir}=8.2\pm0.2\times10^{11}M_{\odot}$.

Another recent M31 mass model by Tamm, Tempel \& Tenjes (2007) and 
Tempel, Tamm \& Tenjes (2007)
also uses Spitzer 3.6 $\mu$m imaging to determine the total mass of M31, and
they find a mass of $1.1\times10^{12} M_{\odot}$ for a pure NFW model, i.e.\
a slightly higher mass than for our best fit. 
However, they did not look at the question
of AC. Also, their bulge seems to have a more concentrated density profile
(see Figure 6 in Tempel et al.\ 2007), which is probably driven by the use
of rotation velocities inside 5 kpc. This may be driving the differences
between our model and the model presented in Tempel et al.\ (2007). We have
been cautious not to include rotation data within the inner few kpc in our
model, since it is unlcear how non-circular motions induced in this barred
region will affect the overall mass model.

It is interesting that, while it has been shown that adiabatically contracted 
halo models do not generally fit the observed rotation curves of galaxies 
(e.g.\ Dutton et al.\ 2005, 2007; Kassin et al.\ 2006a, b; Pizagno et al.\ 
2005), the observed rotation curve of a well-studied galaxy such as M31 can be
well described by a halo that has undergone adiabatic contraction. Indeed, one
either has to adopt {\em (i)} a pure NFW profile with an unphysically large 
concentration, within the context of $\Lambda$CDM, or {\em (ii)} an 
adiabatically contracted NFW profile and a more reasonable concentration. Such
a scenario is in agreement with the results of Klypin et al.\ (2002), who claim
that an adiabatically contracted halo is favoured for M31. 

Much of the power in this analysis came from including the normalization
{\em and} outer slope of M31's rotation curve in our fits (where many similar
studies of AC attempt to fit to just the rotation velocity at 2.2 disc
scalelengths, $V_{2.2}$). This highlights the 
usefulness of extended rotation curve data in constraining general models of 
galaxy formation. Having rotation curve data out to large radii is essential 
for determining dark matter density profiles.

Studies that test the applicability of adiabatic contraction often 
make use of large samples of galaxies that are dominated by late-type 
disc-dominated galaxies, with small bulges or no bulge at all (e.g.\ Pizagno 
et al.\ 2005; Dutton et al.\ 2005, 2007; Gnedin et al.\ 2006; Kassin et al.\ 
2006a, b). These ``bulgeless'' 
galaxies also tend to have very flat (or even 
rising) rotation curves, and it is therefore easy to explain these galaxy 
rotation curves without any need for adiabatic contraction. However, we 
suggest that early-type disc   galaxies (e.g.\ Hubble types Sa--Sb), which 
have 
bulge-dominated centres, and rotation curves that fall-off due to the presence
of a large bulge, may require adiabatic contraction. Although we have studied 
very few galaxies so far (M31 in this paper and two other galaxies in Seigar 
et al.\ 2006) our results indicate that in cases where rotation curves have a
fall-off (two out of three cases), adiabatic contraction is favoured. This may 
be true for all galaxies with large bulges. If so, it may favour a secular 
(rather than merger-driven) origin for these bulge-dominated systems, as the 
gradual accumulation of central mass increases the likelihood that AC will 
operate. 

It should also be noted that in our best-fit model, while it is consistent with
the current H{\tt I} data from Carignan et al.\ (2006), it is unclear if
the rotation curve is declining or remaining flat past 30 kpc, due to the
large error bars. The most important requirement for future kinematical 
studies of galaxies would be improved H{\tt I} measurements at large radii.
This will enable us to put better constraints on the halo properties of M31
and other spiral galaxies.

\section*{Acknowledgments}

Support for this work was provided by NASA through grant number 
HST-AR-10685.01-A from the Space Telescope Science Institute, which is 
operated by the Association of Universities for Research in Astronomy, Inc., 
under NASA contract NAS5-26555. MSS acknowledges partial support from a
Gary McCue Fellowship through the Center for Cosmology at UC Irvine. MSS 
also acknowledges the support of the College of Science and Mathematics
at the University of Arkansas at Little Rock. Research 
by AJB is supported in part by NSF grant AST-0548198. Research by JSB is
supported by NSF grants AST-0507916 and AST-0607377. This work is based in 
part on observations made with the Spitzer Space Telescope, which is operated 
by the Jet Propulsion Laboratory, California Institute of Technology under a 
contract with NASA. This research has made use of the NASA/IPAC Extragalactic 
Database (NED) which is operated by the Jet Propulsion Laboratory, California 
Institute of Technology, under contract with the National Aeronautics and 
Space Administration. The authors wish to thank Tom Jarrett, Luis Ho, Stephane
Courteau and Steve Majewski for their discussions. We also wish to express
gratitude to Pauline Barmby, for supplying us with the most recent mosaic
of the Spitzer 3.6 $\mu$m image of M31. Finally, the authors wish to thank 
the anonymous referees 
for comments which improved the content of this paper.

\appendix

\begin{figure*}
\includegraphics{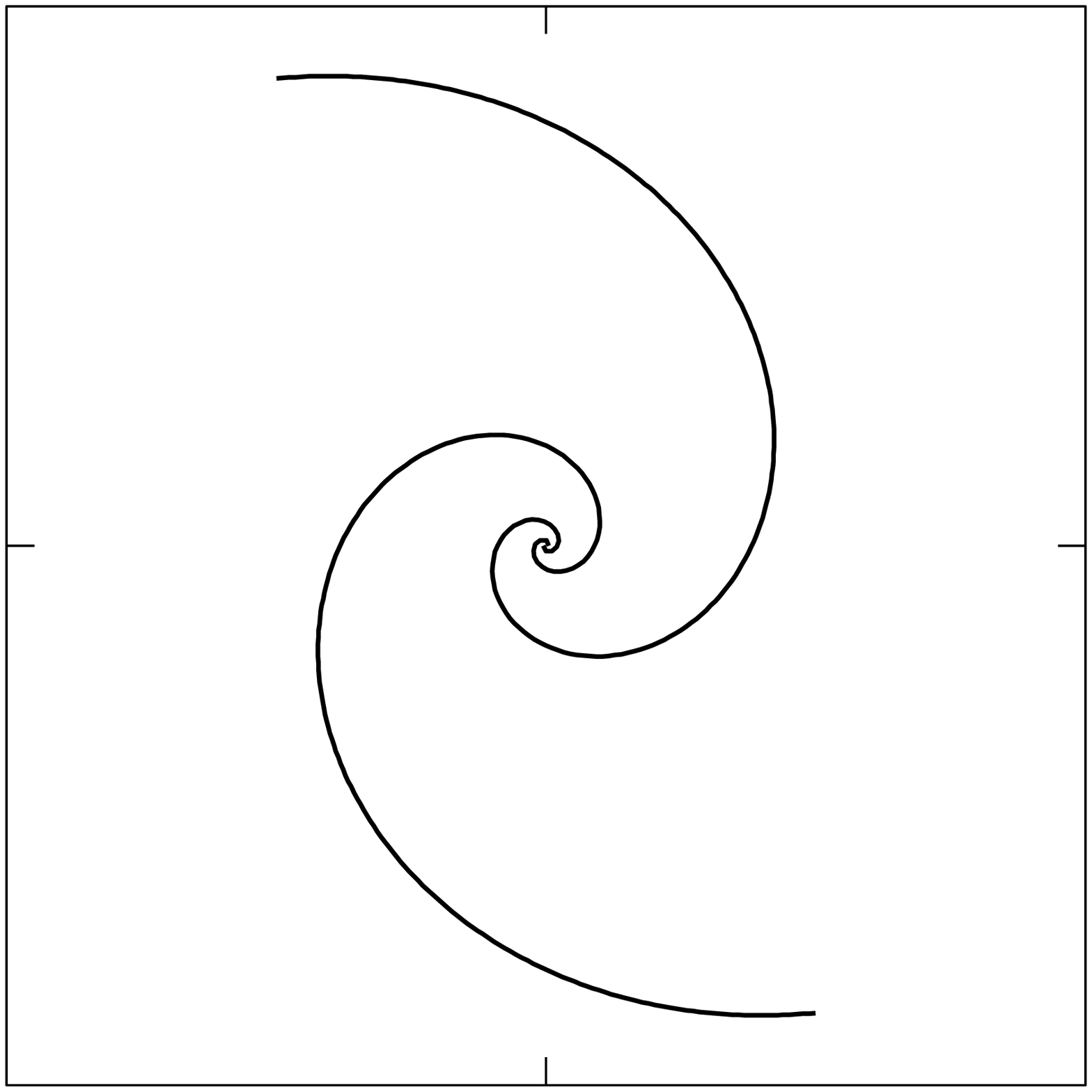}
\includegraphics{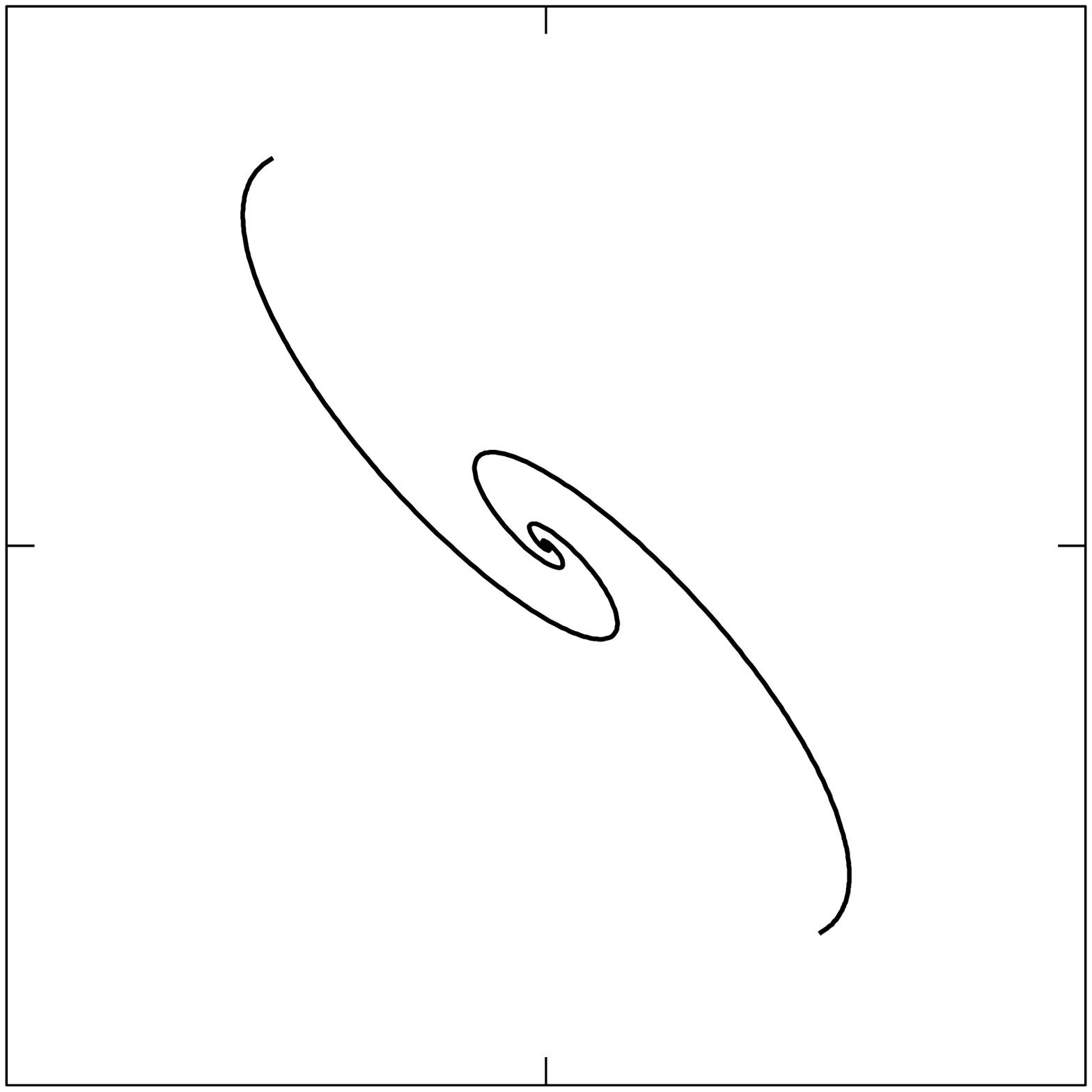}
\includegraphics{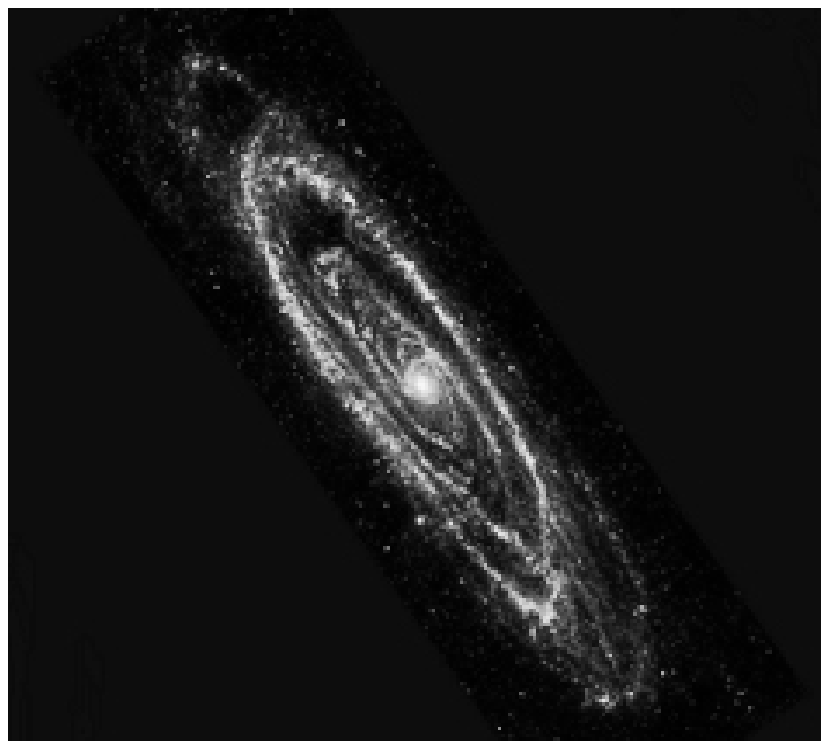}
\vspace*{5.8cm}
\caption{{\em Left:} A two-armed spiral viewed face-on with pitch angle $P=24\fdg7$, as predicted for M31. {\em Centre:} A spiral with pitch angle $P=24\fdg7$, viewed with the inclination and position angle of M31. {\em Right:} 24 $\mu$m Spitzer image of M31 from Gordon et al.\ (2006).}
\label{spiral}
\end{figure*}

\section{The spiral arm pattern of M31}

Because M31 is a highly inclined galaxy, it is difficult to learn how tightly
wound its spiral arm pattern is from imaging data. However, Seigar, Block \&
Puerari (2004) and Seigar et al.\ (2005, 2006) have shown that there is a very
tight correlation between the spiral arm pitch angle (i.e.\ how tightly wound 
the spiral arms are) and rotation curve shear (measured at a 10 kpc radius). 
Therefore the spiral arm pitch angle can be estimated from the shear measured 
for the M31 rotation curve. Rotation curve shear is defined as
\begin{equation}
S=\frac{A}{\omega}=\frac{1}{2}\left(1-\frac{R}{V}\frac{dV}{dR}\right),
\end{equation}
where $A$ is the first Oort constant, $\omega$ is the angular velocity and 
$V$ is the rotational velocity at radius $R$. The shear rate depends upon the
shape of the rotation curve. For a rotation curve that remains flat the shear 
rate, $S=0.5$, for a falling rotation curve the shear rate, $S>0.5$ and for a 
continually rising rotation curve the shear rate, $S<0.5$. Following the 
method described by Seigar et al.\ (2006) we measure the shear at a 10 kpc 
radius. For M31, 10 kpc is equivalent to $43.9^{\prime}$. The shear is 
measured using the prescription described by Seigar (2005) and Seigar et al.\ 
(2004, 2005, 2006). In this method an average slope is fit to the outer part 
of the rotation curve (i.e., past the solid-body rotation regime), and the 
shear is then calculated at the given radius. For M31, at a radius of 10 kpc, 
the shear is $S=A/\omega=0.54\pm0.02$.

There is little scatter in the correlation between 
shear rate and pitch angle, and the line of best fit is given by
\begin{equation}
P=(64.24 \pm 2.87) - (73.24 \pm 5.53)S,
\end{equation}
where $P$ is the spiral arm pitch angle in degrees and $S$ is the shear. From 
this the pitch angle for M31 is $P=24\fdg7\pm4\fdg4$. (It should be noted that
the shear at 10 kpc determined for the M1 B86 model in section 2.2 is $S=0.52$
and for M1 G05 is $S=0.51$. These estimates are consistent with the shear 
measured here. They result in pitch angles of $P=26\fdg2\pm2\fdg2$ and 
$P=26\fdg8\pm2\fdg2$ respectively, both of which are consistent with the 
pitch angle derived above).

Assuming a logarithmic spiral (see e.g., Seigar \& James 1998b for the 
equations that define a logarithmic spiral), the spiral arm pattern of M31 is 
calculated. Figure \ref{spiral} shows the predicted spiral arm pattern of M31 
for a face-on view ({\em left}) and with an inclination angle, $i=77\fdg5$, 
and a position angle $45^{\circ}$ (de Vaucouleurs et al.\ 1991) appropriate 
for M31 ({\em centre}). 
It should be noted that the position and inclination angles are not fitted
parameters, they are taken from the literature.
The spiral arm pitch angle calculated here is larger than the 
H{\tt I} spiral arm pitch angle of 16$^{\circ}$ calculated by Braun (1991), 
although it is possible that the spiral arm pattern in the neutral gas may be 
a few degrees tighter than that observed in the underlying stellar 
distribution. A pitch angle of 16$^{\circ}$ would be a $\sim$2$\sigma$
outlier in the spiral arm pitch angle versus rotation curve shear 
correlation when the shear, $S=0.54$ (see Figure 3 of Seigar et al.\ 2006).
If this point was added to the correlation, it would be one of the larger
outliers, yet the correlation would still be strong.
For comparison the 24 $\mu$m image is shown in the right panel of Figure 
\ref{spiral}.

It is important to note that the spiral arm pattern shown in Figure 
\ref{spiral} does not take into account the disc scale height and how broad 
the arms are. These plots are intended to illustrate show how tightly wound 
the arm pattern would be if viewed from face-on. When viewed close to edge-on 
(as is the case for M31), the scale height of the disc and breadth of the 
spiral arms make it virtually impossible to determine the spiral arm pitch 
angle directly.

\bsp

\label{lastpage}

\end{document}